\documentclass[journal]{IEEEtran}

\usepackage[utf8]{inputenc}
\usepackage[T1]{fontenc}
\usepackage{booktabs}
\usepackage{color, colortbl}
\usepackage{soul}
\soulregister\cite7
\soulregister\ref7
\soulregister\pageref7
\soulregister\url7
\soulregister\textit7
\usepackage{amsmath}
\usepackage{comment}
\usepackage{bm}
\usepackage{amssymb}

\definecolor{lightyellow}{cmyk}{0,0,0.50,0}
\sethlcolor{lightyellow}
\definecolor{yellow}{cmyk}{0,0,0.50,0}

\ifCLASSINFOpdf
   \usepackage[pdftex]{graphicx}
          \DeclareGraphicsExtensions{.pdf,.jpeg,.png}
\else
                  \fi

\usepackage[tight,footnotesize]{subfigure}

\newsavebox\verbtestsetuno
\newsavebox\verbtestsetdue

\begin{document}
\title{DNA-inspired online behavioral modeling\\and its application to spambot detection}

\author{Stefano~Cresci, 
        Roberto~Di~Pietro,
        Marinella~Petrocchi,
        Angelo~Spognardi,
        and~Maurizio~Tesconi
        \thanks{S. Cresci, M. Petrocchi, and M. Tesconi are is with the Institute of Informatics and Telematics, IIT-CNR, Italy.}
\thanks{R. Di Pietro is with Bell Labs, Nokia, Paris-France; University of Padua, Maths Dept., Italy; and the Institute of Informatics and Telematics, IIT-CNR, Italy.}
\thanks{A. Spognardi is with DTU Compute, Technical University of Denmark, Denmark; and the Institute of Informatics and Telematics, IIT-CNR, Italy.}
}

\markboth{IEEE Intelligent Systems}{Cresci \MakeLowercase{\textit{et al.}}: DNA-inspired online
  behavioral modeling and its application to spambot detection}

\maketitle

\begin{abstract}
We propose a strikingly novel, simple, and effective approach to model online user behavior: 
we extract and analyze \textit{digital DNA} sequences from user online
actions and we use Twitter as a benchmark to test our proposal.
We obtain an incisive and compact DNA-inspired
characterization of user actions.
Then, we apply standard DNA analysis techniques to discriminate between genuine and spambot accounts on Twitter. An experimental campaign supports our proposal, showing its effectiveness and viability.
To the best of our knowledge, we are the first ones to identify and adapt DNA-inspired techniques to online user behavioral modeling.
While Twitter spambot detection is a specific use case on a specific social media, our proposed methodology is platform and technology agnostic, hence paving the way for diverse behavioral characterization tasks.
\end{abstract}


\IEEEpeerreviewmaketitle

\makeatletter{}
\section{Online interactions: the mirror of the soul}
\label{sec:intro}
Social media provide internet users with the opportunity 
to interact for a myriads of goals, which range from planning social events to engaging in commercial
transactions.

Modeling and analyzing online user behaviors 
deserves attention for a variety of
reasons. One  is to mine
substantial information regarding events of public interest~\cite{tsakalidis:2015}.
In addition, linking behaviors to a
ground truth in the past leads to predict what will likely happen in
the future when similar behaviors take place~\cite{li:2014}.
Furthermore, online behavioral analysis helps in detecting fictitious and deceitful accounts that may distribute spam or lead to a bias in the public opinion~\cite{huan:2014}.

Current research on online behaviors exploits different techniques, relying, e.g., on social and interaction graphs~\cite{Yang:2013,jiang2016}, textual content~\cite{tsakalidis:2015,DBLP:conf/aaai/HuTL14}, and other complex data representations~\cite{li:2014}.
The definition of a unifying approach is an open challenge.
Thus, we introduce the novel notion of \textit{digital DNA} sequences to
characterize online user behaviors on social media. 
We believe that the high flexibility that is a feature of digital DNA sequences, 
makes this original modeling
technique well suited for different scenarios, with the potential to open up new directions for research.
By drawing a parallel with biological DNA, it also opens up the possibility to draw upon decades of research and development in bioinformatics.

We envisage digital DNA sequences to embrace different applications, among which:
\begin{itemize}
\item letting behavioral patterns emerge from the crowd~\cite{li:2014,jiang2016}, by making use of standard DNA sequence alignment tools.
This automated approach has the reassuring outcome of avoiding the
  often frustrating intervention of humans, who may not
  have the means to discriminate patterns by simply
  inspecting them on an ``account-by-account'' basis;
 \item defining a behavioral-based
   taxonomy of online interactions. For social media, this might mean classifying users as compulsive,
   curious, lazy, inactive. Such information could then be exploited to
   launch ad-hoc marketing campaigns and to convey specific messages
   to those accounts that are more likely to accept such suggestions.
\end{itemize}

\makeatletter{}
\section{Concept and instances of Digital DNA}
\label{sec:genome}
\subsection{Biological DNA} 
The human genome is the complete set of genetic information on humans (\textit{Homo sapiens}), encoded as nucleic acid (DNA) sequences. DNA sequences are successions of characters (i.e., strings), indicating the order of nucleotides within DNA molecules. The  possible characters are \textbf{A}, \textbf{C}, \textbf{G}, and \textbf{T}, representing the four nucleotide bases of a DNA strand: adenine, cytosine, guanine, thymine.
Biological DNA stores the information directing functions and characteristics of a living organism. DNA sequences are analyzed by means of bioinformatics techniques. Sequence alignment and motif elicitation are among the most well-known analysis techniques to find sequence commonalities and repetitions. Via an analysis of common sub-sequences, it is possible to predict specific characteristics of the individual and to uncover relationships between different individuals.

\subsection{Digital DNA as a proxy of user online behavior}
Inspired by biological DNA, we propose to model online user behaviors with strings of characters representing the sequence of a user's online actions. Each action type (e.g., posting new content, following or replying to a user) can be encoded with a different character, similarly to DNA sequences, where 
characters encode nucleotide bases. According to this paradigm, online user actions represent the bases of their \textit{digital DNA}.

Different kinds of user behaviors can be observed on the Internet~\cite{DBLP:conf/aaai/HuTL14} and digital DNA is a flexible and compact way of modeling such behaviors.
Its flexibility lies in the possibility to 
choose which actions will form the sequence. 
For example, digital DNA sequences on Facebook
could include a different base for each user-to-user interaction type: comments (\textbf{C}), likes (\textbf{L}), shares (\textbf{S}) and mentions (\textbf{M}).
Then, interactions are encoded as strings formed by such characters, according to the sequence of actions users perform.
Similarly, user-to-item interactions on an e-commerce platform could be modeled using a base for every product category. User purchasing behaviors could be encoded as a sequence of characters, according to the category of products they buy.
To this regard, digital DNA shows a major difference to biological DNA, where the four nucleotide bases are fixed. In digital DNA, both the number and the meaning of the bases can change according to the behavior/interaction to be modeled.
Similarly to its biological predecessor, digital DNA is a compact representation of information. For example, the timeline of a Twitter user could be encoded as a single string of 3,200 characters (one character per tweet). There is a vast number of algorithms and techniques to draw upon for the analysis of digital DNA sequences. Indeed, many of the techniques developed 
in the field of bioinformatics for the analysis of biological DNA can be carried over to study the characteristics of digital DNA as well.

In the following, we show the
usefulness of our approach with a possible application to the
detection of Twitter spambots, namely computer programs that control
social accounts with the goal to mimic real users and send these users unsolicited messages~\cite{DBLP:conf/aaai/HuTL14}.
Starting from two Twitter datasets where
genuine and spambot accounts are a priori known, we leverage DNA sequence characterization to let recurrent
patterns emerge.
We show how groups of spambots share common patterns, as opposed to groups of
genuine accounts. As a concrete application of this outcome, we
demonstrate how to apply the methodology to tell apart spambots from
genuine accounts, within an unknown set of accounts.

\makeatletter{}
\section{Introducing digital DNA fingerprinting}
\label{sec:fingerprinting}

Biological DNA fingerprinting is a technique employed by forensic scientists to identify suspects from their DNA.
Similarly to forensic analysis, we describe a digital DNA fingerprinting technique for spambots detection on social media.
While this is just one of the possible applications for digital DNA, it contributes to understand how our methodology can help in practice. 

Academics and platform administrators are constantly struggling to keep pace with evolving spambots. New waves of malicious accounts present different and advanced features, making their detection with existing systems extremely challenging~\cite{Yang:2013,DBLP:conf/aaai/HuTL14}.

This is the case with the new family of spambots that emerged on Twitter during the last Mayoral election in Rome, 2014. One of the runners-up made use of almost 1,000 automated accounts for his electoral campaign in order to pubblicize his policies.
Surprisingly, these automated accounts were extremely hard to distinguish from genuine ones. Each profile contained detailed personal information and had thousands of genuine followers and friends.
Furthermore, they showed a tweeting behavior apparently similar to that of genuine accounts, with a few tweets posted every day -- mainly quotes from popular people. However, every time the political candidate posted a new tweet from his official account, all the spambots retweeted it in a time span of just a few minutes. 
Thus, the candidate was able to reach many more accounts than his own direct followers and managed to alter Twitter engagement metrics during the electoral campaign.
We found similar phenomena to be widespread also outside Italy. Specifically, here we consider two groups of social spambots: (i) bot retweeters of the Italian candidate (\verb|Bot1|), and (ii) bot accounts spamming URLs pointing to several products on sale on the \textit{Amazon.com} e-commerce platform (\verb|Bot2|).

In contrast with classification and supervised approaches, we devise an unsupervised way to detect spambots, by comparing their behavior with the aim of finding similarities between automated accounts. We model the behavior of the two groups of spambots via their digital DNA and we compare it to that of a sample of genuine accounts.  
We exploit digital DNA to study the behavior of groups of users following the intuition that, because of their automated nature, spambots are likely to present higher similarities in their digital DNA with respect to the more heterogeneous genuine users.

\subsection{Mining groups of digital DNA sequences}
The basic assumption of the digital DNA fingerprinting technique is to consider behaviors as sequences of actions and to characterize and detect social spambots by grouping similar sequences.
Our approach thus falls within the broad field of sequential data mining, yet it presents differences from the tasks commonly performed when working with sequential data.
Sequences are ordered lists of symbols from an alphabet
$\mathbb{B}$, namely $\bm{s}=\{s_1,s_2,\ldots,s_n\}$, with
$s_i\in\mathbb{B}$ and are used in information retrieval,
part-of-speech tagging, genomics, time-series analysis -- to name but a few. Tasks commonly performed with such sequences are those of sequential supervised learning, time-series prediction, sequence classification.
For instance, in sequential supervised learning, given a sequence $\bm{s}=\{s_1,s_2,\ldots,s_n\}$, the goal is to find a sequence of labels $\bm{l}=\{l_1,l_2,\ldots,l_n\}$, where each label is associated to an element of the original sequence.
Well-known techniques to perform this task are conditional random fields and hidden Markov models.
Instead, in time series prediction, the goal is to find the element $s_{t+1}$, given a starting sequence $\bm{s}=\{s_1,s_2,\ldots,s_t\}$. For instance, this goal could be achieved by employing a model taken from the broad family of statistical (auto-)regressive techniques.
Finally, sequence classification is concerned with assigning a single label $l$ to a whole sequence $\bm{s}=\{s_1,s_2,\ldots,s_n\}$, following a supervised approach. 
Instead, here our goal is to analyze a set of unlabeled sequences, in order to find commonalities and differences between them, 
by exploiting the sequential information related to user actions, which is captured within digital DNA sequences. To this regard, our task resembles that of unsupervised sequence clustering. However, it further differs from traditional machine learning clustering since we exploit sequential data. Indeed, our sequences are ordered data representations, while feature vectors exploited in traditional clustering are unordered (i.e., features are not ordered in feature vectors and the ordering is not exploited in the analysis).
Moreover, our sequences are variable-length vectors of symbols drawn from a limited alphabet (i.e., strings), rather than fixed-length numeric vectors. Thus, traditional distance/similarity metrics for numeric vectors (such as those typically used in clustering) are not applicable in our case.
Working with ordered and variable-length data representations also marks a difference with other techniques recently employed for the behavioral analysis of groups of users, such as those based on hashing~\cite{ou2015} and graph mining~\cite{jiang2016}.

\makeatletter{}
\section{Digital DNA fingerprinting\\for spambot detection}
\label{sec:application}
\subsection{Characterizing spambot behavior}
\label{sec:characterizing}
The process of digital DNA fingerprinting has four main steps: (i) acquisition of behavioral data; (ii) extraction of DNA sequences; (iii) comparison of DNA
sequences; (iv) evaluation. 

In order to perform a rigorous evaluation of our detection technique, we created datasets of verified spambots and genuine Twitter accounts.
All the accounts of our datasets have undergone manual verification. Specifically, 50.05\% (991 accounts) of \verb|Bot1| accounts and 89.29\% (464 accounts) of \verb|Bot2| accounts were certified as spambots.
Then, we built a dataset of human accounts by randomly contacting Twitter users and by asking them simple questions in natural language. All the 3,474 accounts that answered were certified as human.
For all the 4,929 accounts of our datasets we then collected behavioral data, by crawling the content of their Twitter timelines.

\begin{figure*}
  \centering
  \subfigure[Tweet type DNA similarity between retweeters of the political candidate and genuine accounts.
  \label{fig:comparison-exp1-bot1}]{\includegraphics[width=0.42\textwidth]{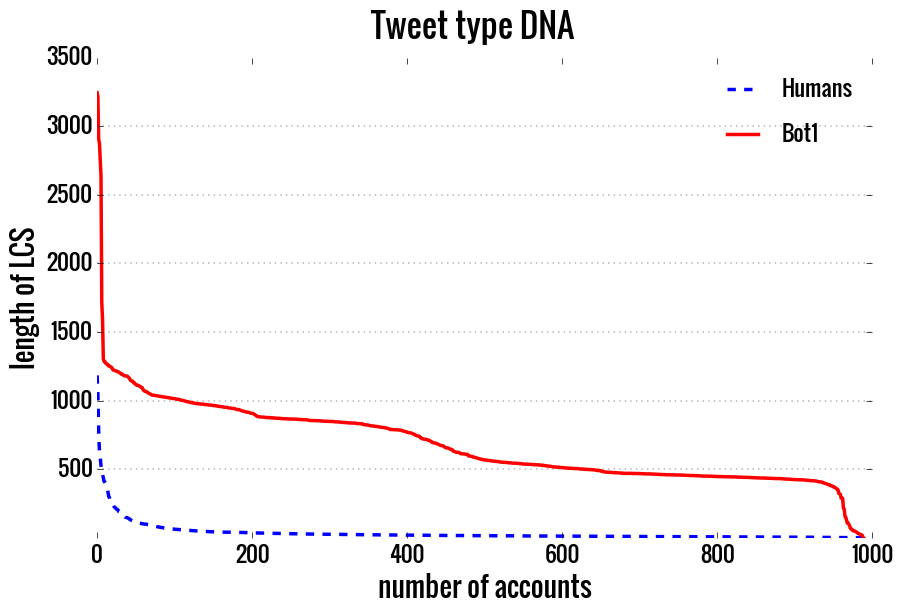}}
  \subfigure[Tweet type DNA similarity between spammers of Amazon.com products and genuine accounts.
  \label{fig:comparison-exp1-bot2}]{\includegraphics[width=0.42\textwidth]{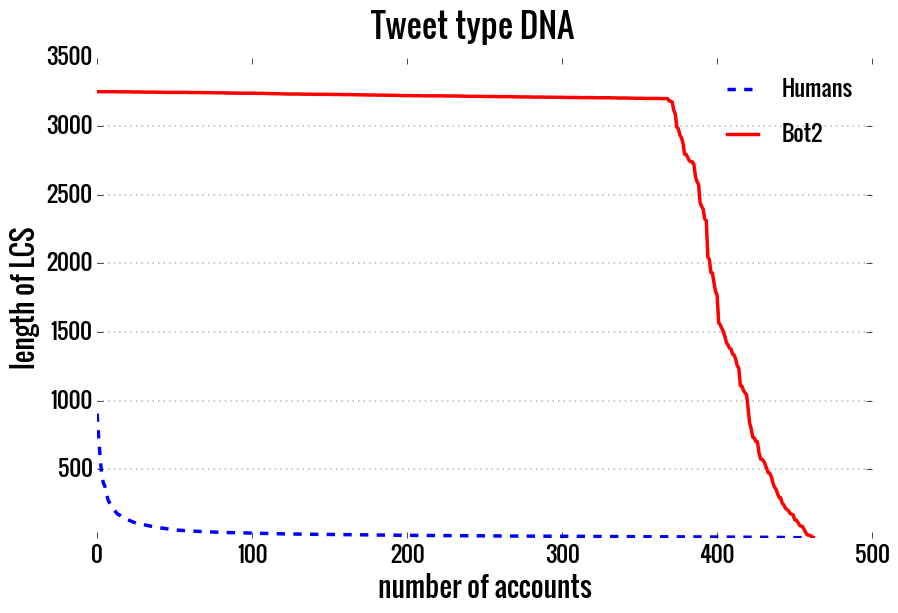}} \\
  \subfigure[Tweet content DNA similarity between retweeters of the political candidate and genuine accounts.
  \label{fig:comparison-exp2-bot1}]{\includegraphics[width=0.42\textwidth]{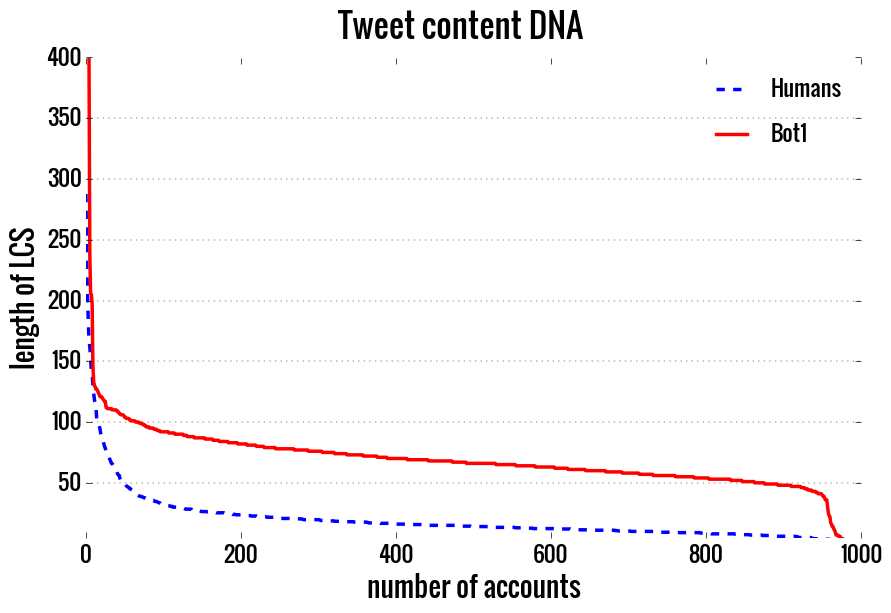}}
  \subfigure[Tweet content DNA similarity between spammers of Amazon.com products and genuine accounts.
  \label{fig:comparison-exp2-bot2}]{\includegraphics[width=0.42\textwidth]{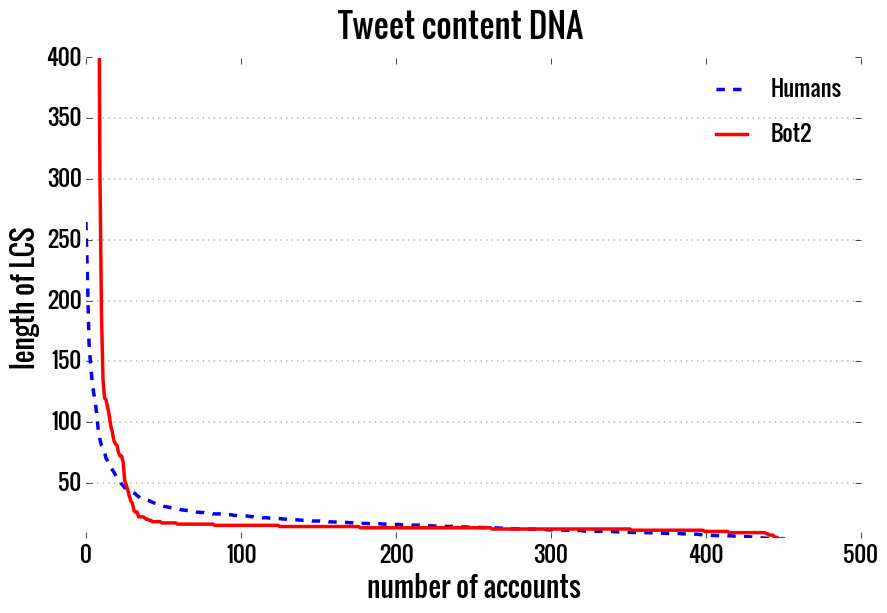}}
  \caption{Comparison between the LCS lengths of the digital DNA of different groups of Twitter accounts.
  \label{fig:comparison}}
\end{figure*}

The second step is the extraction of the digital DNA of the accounts -- that is, associating each account to a string that encodes its behavioral information.
Digital DNA sequences can be extracted in different ways, according to the kind of information to be modeled.
Similarly to feature engineering in machine learning, catching the right granularity for modeling behaviors might lead to better detection results.
For the sake of experimentation, we modeled the accounts' behavior in two ways and we evaluated which one better captures the nature of automated accounts.
In the first experiment, we considered the \textit{types} of tweets shared (tweet type DNA). Every tweet in the account's timeline is encoded with a different character: \textbf{A} for a simple tweet, \textbf{T} for a reply and \textbf{C} for a retweet. In the second experiment, we considered the \textit{content} of tweets (tweet content DNA). We designed the base \textbf{A} for tweets with URLs, \textbf{T} for tweets with hashtags, \textbf{C} for tweets with mentions, \textbf{G} for tweets with media (pictures, videos), \textbf{X} for tweets with a combination of the previous entities, and \textbf{N} for tweets with none of them (plain-text). Since each Twitter timeline is made up of at most 3,200 tweets, we obtained digital DNA sequences counting up to 3,200 characters.

\begin{lrbox}{\verbtestsetuno}
\texttt{Test-set1}
\end{lrbox}
\begin{lrbox}{\verbtestsetdue}
\texttt{Test-set2}
\end{lrbox}

\begin{figure*}
  \centering
  \subfigure[DNA similarity of a mixed group of genuine accounts and retweeters of the political candidate (\texttt{Test-set1}). \label{fig:testset-bot1}]{
  \includegraphics[width=0.42\textwidth]{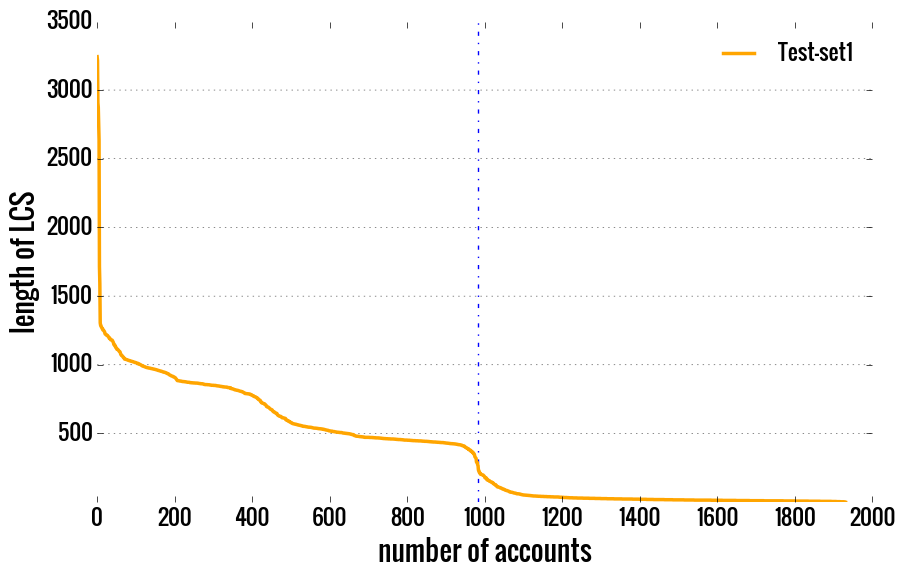}}
  \subfigure[DNA similarity of a mixed group of genuine accounts and spammers of Amazon.com products (\texttt{Test-set2}). \label{fig:testset-bot2}]{
  \includegraphics[width=0.42\textwidth]{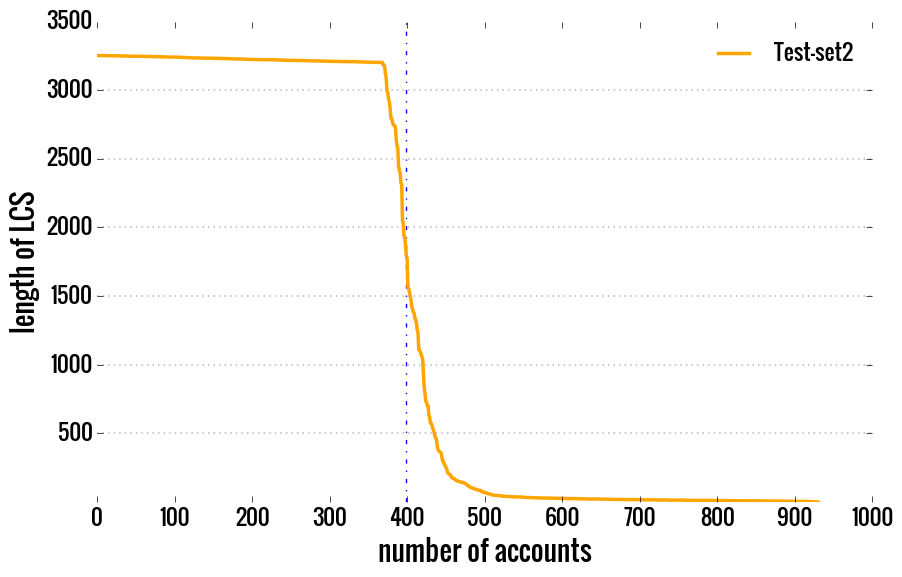}} \\
  \subfigure[Evaluation of spambot detection results for a mixed group of genuine accounts and retweeters of the political candidate (\texttt{Test-set1}). \label{tab:evaluation-testset1}]{
  \makeatletter{}
{\small
\begin{tabular}{lc@{\phantom{M}}rrrrc@{\phantom{M}}rrrrrr}
	\toprule
	&& \multicolumn{4}{c}{\textbf{detection results}} && \multicolumn{6}{c}{\textbf{evaluation metrics}} \\
	\cmidrule{3-6} \cmidrule{8-13}
	\textbf{detection system} && \textit{TP} & \textit{TN} & \textit{FP} & \textit{FN} && \textit{Precision} & \textit{Recall} & \textit{Specificity} & \textit{Accuracy} & \textit{F-Measure} & \textit{MCC}\\
	\midrule
					Yang \textit{et al.}~\cite{Yang:2013}		&& 169	& 811	& 131	& 822	&& 0.563	& 0.170	& 0.860	& 0.506	& 0.261	& 0.043 \\ [1.25ex]
					Miller \textit{et al.}~\cite{miller2014}		&& 355 	& 657	& 285	& 636	&& 0.555	& 0.358	& 0.698	& 0.526	& 0.435	& 0.059 \\ [1.25ex]
					Ahmed \textit{et al.}~\cite{ahmed2013}	&& 935	& 888	& 54		& 56		&& 0.945	& 0.944	& 0.945	& 0.943	& 0.944	& 0.886 \\ [1.25ex]
					DNA fingerprinting					&& 963	& 924	& 18		& 28		&& 0.982	& 0.972	& 0.981	& 0.976	& 0.977	& 0.952 \\
	\bottomrule
\end{tabular}}

 } \\
  \subfigure[Evaluation of spambot detection results for a mixed group of genuine accounts and spammers of Amazon.com products (\texttt{Test-set2}). \mbox{$^{\sharp}$: With regards} to the feature set of~\cite{ahmed2013}, a few accounts had null values for all the features thus resulting in the impossibility to apply the clustering algorithm to such accounts. \label{tab:evaluation-testset2}]{
  \makeatletter{}
{\small
\begin{tabular}{lc@{\phantom{M}}rrrrc@{\phantom{M}}rrrrrr}
	\toprule
	&& \multicolumn{4}{c}{\textbf{detection results}} && \multicolumn{6}{c}{\textbf{evaluation metrics}} \\
	\cmidrule{3-6} \cmidrule{8-13}
	\textbf{detection system} && \textit{TP} & \textit{TN} & \textit{FP} & \textit{FN} && \textit{Precision} & \textit{Recall} & \textit{Specificity} & \textit{Accuracy} & \textit{F-Measure} & \textit{MCC}\\
	\midrule
					Yang \textit{et al.}~\cite{Yang:2013}		&& 190	& 397	& 71		& 274	&& 0.727	& 0.409	& 0.848	& 0.629	& 0.524	& 0.287 \\ [1.25ex]
					Miller \textit{et al.}~\cite{miller2014}		&& 142	& 306	& 162	& 322	&& 0.467	& 0.306	& 0.654	& 0.481	& 0.370	& -0.043 \\ [1.25ex]
					Ahmed \textit{et al.}~\cite{ahmed2013} $^{\sharp}$	&& 428	& 427	& 41		& 30		&& 0.913	& 0.935	& 0.912	& 0.923	& 0.923	& 0.847 \\ [1.25ex]
					DNA fingerprinting					&& 398	& 468	& 0		& 66		&& 1.000	& 0.858	& 1.000	& 0.929	& 0.923	& 0.867 \\
	\bottomrule
\end{tabular}}

 }
  \caption{DNA similarity of the two test-sets and evaluation of spambot detection results.
  \label{fig:testset}}
\end{figure*}

As a third step, we study similarities among the DNA of our accounts.
We consider similarity as a proxy for automation and, thus, an exceptionally high level of similarity among a large group of accounts serves as a red flag for anomalous behaviors. 
We quantify similarity by looking at the Longest Common Substring (LCS) among digital DNA sequences.
We rely on a linear time algorithm~\cite{arnold:2009} that is based on the generalized suffix tree. Given $m$ strings, the algorithm is able to find, for all
$2\leq k\leq m$, the longest substring in common to at least $k$
strings. Then, given $m$ DNA sequences and the length of the
longest substring common to $k$ of them, we can derive which
accounts have such longest substring in common (i.e., which are those $k$ accounts).
As the number $k$ of accounts grows, the length of the longest substring common to all of them (LCS) shortens.
Thus, it is likely to find a long LCS among a few accounts rather than among large groups. 
Thus, the rationale behind our analysis is that if the LCS is long when $k$ grows, then the accounts that share that LCS
have a suspiciously similar behavior.

To verify our claim, we computed the LCS in the datasets of certified spambots and genuine accounts for every $k$.
To be statistically sound, we compared groups of the same size by randomly undersampling genuine accounts to match the number of spambots.
Figures~\ref{fig:comparison-exp1-bot1} and~\ref{fig:comparison-exp1-bot2} show results of the comparison of tweet type DNA sequences, while figures~\ref{fig:comparison-exp2-bot1} and~\ref{fig:comparison-exp2-bot2} show the results for tweet content DNA sequences.
All plots of Figure~\ref{fig:comparison} show the length of LCS for all the possible group sizes: for example, in Figure~\ref{fig:comparison-exp1-bot1} there exists a
group of 400 accounts in \verb|Bot1| with a substring of length 774 in common.
From figures~\ref{fig:comparison-exp1-bot1} and~\ref{fig:comparison-exp1-bot2} it is clear that the LCS of both the groups of spambots
are rather long even when the number of accounts grows. This is strikingly evident for
\verb|Bot2| (spammers of \textit{Amazon.com} products), see Figure~\ref{fig:comparison-exp1-bot2}.
In contrast, genuine accounts show little to no similarity.
For both the spambot groups we observe a sudden drop in LCS length when $k$ 
gets close to the group size. Instead, the human group has an exponential decay and it rapidly reaches the smallest values of the LCS length.

Figures~\ref{fig:comparison-exp2-bot1} and~\ref{fig:comparison-exp2-bot2} are related to tweet content DNA sequences and results are noticeably different from those obtained with tweet type DNA.
All the LCS curves have much lower values, however the biggest difference between tweet content DNA and tweet type DNA is shown for \verb|Bot2|. While tweet type DNA models almost perfectly the high similarity of \verb|Bot2| accounts, tweet content DNA is unable to show a significative difference between genuine and bot accounts.
These results suggest that modeling the behavior of Twitter accounts according to the type of tweets, rather than to their content, is more effective for discriminating \verb|Bot2| spambots.

\subsection{Uncovering novel Twitter spambots}\label{subsec:uncovering}
As shown before, properly extracted digital DNA sequences  provide evidence of similar activities in large groups.
In turn, this can lead to suspect a high degree of automation for those accounts behaving in a similar way.
Building on the promising results obtained with tweet type DNA, in the following we exploit the characteristics of LCS curves to effectively detect groups of Twitter spambots. Our detection technique can be applied to a group of unlabeled accounts to check whether accounts with a suspiciously similar behavior are present.
In a group with mixed bot and genuine accounts, almost only bot accounts will have long DNA substrings in common.
Thus, by identifying the group of accounts that share a long LCS, we can obtain a set of suspicious accounts.

Henceforth, \verb|Test-set1| and \verb|Test-set2| refer to unlabeled groups where genuine accounts have been mixed with the ones from \verb|Bot1| and \verb|Bot2| respectively.

The plots in figures~\ref{fig:testset-bot1} and~\ref{fig:testset-bot2} show the length of LCS for all the possible values of $k$ accounts in \verb|Test-set1| and \verb|Test-set2|. 
In these plots, we looked for the points where the LCS curves exhibit a sudden drop to very low values, since those points might represent thresholds between groups of similar accounts. 
Observing Figure~\ref{fig:testset-bot1}, we can see that there is a continuous decrease in the LCS length as the number of accounts grows, and a sudden drop occurs just before reaching the 1,000 accounts. The drop is even more evident in Figure~\ref{fig:testset-bot2} where there is a steep fall of the curve just before 400 accounts. LCS curves in both plots keep approaching to 1 as the number of accounts grows. 
The steep drops of LCS curves highlight areas where the length of the LCS remains practically unchanged even for significantly different numbers of accounts considered. In fact, such \textit{plateaux} in LCS curves are strictly related to homogeneous groups of highly similar accounts. Note that it is possible to observe multiple plateaux in a single LCS curve, as in the case of Figure~{\ref{fig:testset-bot1}}. This represents a situation where multiple (sub-)groups exist among the whole set of considered accounts. Furthermore, the steeper and the more pronounced the drop in the LCS curve, the more different are the two groups of accounts split by that drop.

Building on these interesting characteristics, we devised a methodology to exploit drops in LCS curves in order to identify groups of similar accounts. Specifically, we exploit the derivative of LCS curves $\frac{\Delta\;\text{LCS}}{\Delta\;\text{accounts}}$ to highlight the points corresponding to the steepest drops. Such points, appearing as negative peaks in the derivative plot, represent good candidate splitting points to detect groups of dissimilar accounts. Note that it is possible to obtain several candidate splitting points, which might also be ranked according to the derivative value (i.e., how steep is the corresponding drop). Then, a hierarchical top-down (i.e., divisive) approach may be applied by repeatedly splitting the set of accounts based on the ranked candidate points, leading to a dendrogram-like structure. For instance, this approach can be exploited in situations where the LCS curve exhibits multiple plateaux and steep drops.

Thus, analyzing the LCS curves of figures~{\ref{fig:testset-bot1}} and~{\ref{fig:testset-bot2}}, we identified the best splitting points and we assumed that the suspiciously similar accounts are the ones that share the longest LCS just before such splitting points, namely the 983 accounts with LCS around 400 for \texttt{Test-set1} and the 400 accounts with LCS of around 1750 for \texttt{Test-set2}.
Finally, we evaluated the effectiveness of the DNA-based detection technique by considering the following evaluation metrics: Precision, Recall, Specificity, Accuracy, F-Measure, and Matthews Correlation Coefficient (MCC).
To thoroughly evaluate the DNA fingerprinting technique we compared our detection results with those obtained by several different state-of-the-art approaches, namely the supervised one by Yang \textit{et al.}~{\cite{Yang:2013}} and the unsupervised approaches by Miller \textit{et al.}~{\cite{miller2014}} and by Ahmed \textit{et al.}~{\cite{ahmed2013}}.
The work presented in~\cite{Yang:2013} provides a machine learning classifier that infers whether a Twitter account is genuine or spambot by relying on account's relationships, tweeting timing and level of automation.
We have reproduced such classifier since the authors kindly provided us their training set. Instead, works in~{\cite{miller2014}} and~{\cite{ahmed2013}} define a set of machine learning features and apply clustering algorithms. Specifically, in~{\cite{miller2014}} the authors propose modified versions of the DenStream and StreamKM++ algorithms (respectively based on DBSCAN and k-means) and apply them for the detection of spambots over the Twitter stream. Ahmed \textit{et al.}~{\cite{ahmed2013}} exploit the Euclidean distance between feature vectors to build a similarity graph of the accounts. Graph clustering and community detection algorithms are then applied to identify groups of similar accounts in the graph.

Tables~{\ref{tab:evaluation-testset1}} and~{\ref{tab:evaluation-testset2}} show the results of this comparison. 
Notably, the DNA fingerprinting detection technique outperforms all other approaches, achieving $\textit{MCC} = 0.952$ for \texttt{Test-set1} and $\textit{MCC} = 0.867$ for \texttt{Test-set2}.
Specifically, there is a clear performance gap between the approaches of~{\cite{Yang:2013}} and~{\cite{miller2014}} with respect to our proposed approach and that of~{\cite{ahmed2013}}. The supervised approach by Yang \textit{et al.}~{\cite{Yang:2013}} proved unable to accurately distinguish spambots from genuine accounts, as demonstrated by the considerable number of false negatives (\textit{FN}) and the resulting very low \textit{Recall}.
This result supports our initial claim that this new wave of bots is surprisingly similar to genuine accounts: they are exceptionally hard to detect if considered one by one.
Moreover, also the unsupervised approach in~{\cite{miller2014}} provided unsatisfactory results. Among the 126 features proposed in~{\cite{miller2014}}, 95 are based on the textual content of tweets. However, novel social spambots
tweet contents similar to that of genuine accounts (e.g., retweets of genuine tweets and famous quotes).
Instead, the approach in~{\cite{ahmed2013}} proved effective in detecting our considered spambots, showing an $\textit{MCC} = 0.886$ for \texttt{Test-set1} and $\textit{MCC} = 0.847$ for \texttt{Test-set2}. Employing only 7 features, {\cite{ahmed2013}} focuses on retweets, hashtags, mentions, and URLs, thus analyzing the accounts along the dimensions exploited by these spammers. However, although achieving an overall good performance for the considered spambots, the approach in~{\cite{ahmed2013}} might lack reusability across other groups of spambots with different behaviors, such as those
perpetrating a follower fraud~\cite{jiang2016},\cite{cresci2015}.
Instead, our DNA-inspired technique is flexible enough to highlight suspicious similarities among groups of accounts without focusing on specific characteristics.

Finally, it is worth noting that some of the state-of-the-art approaches for spambots detection require a large number of data-demanding features. For instance, approaches that are based on graph mining have been proved to be more demanding in terms of data that is needed in order to perform the detection~{\cite{cresci2015}}. Instead, the DNA fingerprinting technique only exploits Twitter timeline data to perform spambots detection.
Furthermore, it does not require a training phase and can be employed pretty much like a clustering algorithm, in an unsupervised fashion.

We also envision the possibility to exploit results of our DNA-inspired technique as a feature in a more complex detection system. Indeed, different types of DNA (such as the tweet type DNA and the tweet content DNA) can be exploited to model user behaviors along different directions. Then, results of these models could be used simultaneously in an ensemble or voting system. To this regard, the already interesting results achieved by exploiting only one type of digital DNA, namely tweet type DNA, might represent promising ground for further experimentation and research.

\section{Conclusion}
In this paper we have introduced a novel methodology to characterize and study online user behaviors.
The novelty of our proposal relies in modeling behaviors in a DNA-inspired fashion.
Hence, we enable analysts to leverage a powerful set of tools -- that have been developed over decades for DNA analysis -- to validate their working hypotheses on online user behaviors. 
To show the viability of our proposal, we have performed an extensive experimental campaign using Twitter as a benchmark and targeting spambots detection.
Results show that our proposal outperforms best-of-breed algorithms that are commonly employed for this task.
Finally, we also argue that the proposed methodology could be used in other contexts as well, hence paving the way for further research.

\ifCLASSOPTIONcaptionsoff
  \newpage
\fi

\bibliographystyle{IEEEtran}
\bibliography{manuscript}

\begin{IEEEbiographynophoto}{Stefano Cresci}
is a Ph.D. student at University of Pisa, Information Engineering Dept., and a research fellow at IIT-CNR in Pisa, Italy. His research
  interests include social media mining and knowledge discovery.\\
  Email: s.cresci@iit.cnr.it
\end{IEEEbiographynophoto}
\begin{IEEEbiographynophoto}{Dr.~Roberto Di Pietro}
is with the Nokia Bell Labs, Paris, France.
His research interests include security, privacy, distributed systems, computer forensics, and analytics.
He is also affiliated with University of Padua, Maths Dept., and IIT-CNR, Pisa, Italy.\\
  Email: roberto.di\_pietro@nokia.com
\end{IEEEbiographynophoto}
\begin{IEEEbiographynophoto}{Dr.~Marinella Petrocchi} is a researcher at
  IIT-CNR, Pisa, Italy. 
  Her research interests include data privacy, 
 personalization on the internet, data quality and trustworthiness.\\
  Email:  m.petrocchi@iit.cnr.it
\end{IEEEbiographynophoto}
\begin{IEEEbiographynophoto}{Dr.~Angelo Spognardi}
  is an assistant professor at DTU Compute, Denmark. 
  His main research interests are on social
  networks modeling and analysis,
  network protocol security and privacy.\\
  Email: angsp@dtu.dk
\end{IEEEbiographynophoto}
\begin{IEEEbiographynophoto}{Dr.~Maurizio Tesconi}
  is a researcher at IIT-CNR in Pisa, Italy. 
  His research interests cover
  semantic web, social network analysis and mining, 
  visual analytics, lexical resources technologies
  and paperless solutions for the public
  administration sector.\\
  Email: m.tesconi@iit.cnr.it
\end{IEEEbiographynophoto}

\end{document}